# Quantum Algorithms of Solving the Backtracking of One-dimensional Cellular Automata

Weng-Long Chang[1]

[1]Contact Author: Department of Computer Science and Information Engineering, National Kaohsiung University of Applied Sciences, Kaohsiung City, Taiwan 807-78, Republic of China
E-mail: changwl@cc.kuas.edu.tw

Mang Feng[2]

[2]State Key Laboratory of Magnetic Resonance and Atomic and Molecular Physics, Wuhan Institute of Physics and Mathematics, Chinese Academy of Sciences, Wuhan, 430071, People's Republic of China
E-mail: mangfeng@wipm.ac.cn

Kawuu Weicheng Lin[3]

[3]Department of Computer Science and Information Engineering, National Kaohsiung University of Applied Sciences, Kaohsiung City, Taiwan 807-78, Republic of China
E-mail: linwc@cc.kuas.edu.tw

Chih-Chiang Wang[4]

[4]Department of Computer Science and Information Engineering, National Kaohsiung University of Applied Sciences, Kaohsiung City, Taiwan 807-78, Republic of China
E-mail: Steven.cc.wang@gmail.com

Ju-Chin Chen[5]

[5]Department of Computer Science and Information Engineering, National Kaohsiung University of Applied Sciences, Kaohsiung City, Taiwan 807-78, Republic of China
E-mail: joan@csie.ncku.edu.tw

_________________________________________________________________________________________

In [Wolfram 1982; Wolfram 1983; Wolfram 2002], the backtracking of one-dimensional cellular automata is to find out which of the $2^n$ possible initial configurations of width $n$ evolve to a specific configuration. In this paper, in one-dimensional cellular automata for a specific configuration of width $n$, its unique initial configuration can be found by mean of the proposed quantum algorithm with polynomial quantum gates, polynomial quantum bits and the successful probability that is the same as that of Shor's quantum order-finding algorithm in [Shor 1994].

_________________________________________________________________________________________

## 1. INTRODUCTION

Cellular automata were proposed by von Neumann and Ulam in [Neumann and Ulam 1966]. From [Wolfram 1982; Wolfram 1983], any system with many identical discrete elements undergoing deterministic local interactions may be modeled as cellular automaton. For example, from [Neumann and Ulam 1966; Wolfram 1982; Wolfram 1983], self-reproduction in biological processes is cellular automata, and a parallel-processing computer in which the initial configuration encodes the program and input data, and time evolution yields the final output is also cellular automata. It is shown from [Adleman 1994] that a molecular computer with biological operations and DNA sequences is cellular automata in nature. In [Wolfram 2002], cellular automata are designed to deal with relative questions of mathematics, physics, biology, social sciences, computer science, philosophy and art. One of the interesting open questions is to ask how to solve the backtracking of one-dimensional cellular automata in time.

In [Wolfram 1982; Wolfram 1983; Wolfram 2002], the backtracking of one-dimensional cellular automata is to find out which of the $2^n$ initial configurations of width $n$ evolve to a specific configuration. In this paper, it is shown that for given one specific configuration of width $n$ in one-dimensional cellular automata, the quantum circuit of implementing the evolved processing for the $2^n$ initial configurations of width $n$ is responsible for processing all the computational basis states and labeling the corresponding unique initial configurations of width $n$. Next, it is demonstrated that amplitude amplification of the corresponding unique initial configurations of width $n$ for given one specific configuration of width $n$ in one-dimensional cellular automata can be completed by means of Shor's quantum order-finding algorithm in [Shor 1994]. Then, it is also proved that after a measurement on the answer is completed; the successful probability of obtaining the answer is the same as that of Shor's quantum order-finding algorithm. Furthermore, it is shown that the time complexity and the space complexity of finding out which of the $2^n$ initial configurations of width $n$ evolve to a specific configuration in one-dimensional cellular automata are, respectively, polynomial quantum gates and polynomial quantum bits.

The rest of the paper is organized as follows: in Section 2, abstracted problems are introduced and the motivation to develop quantum algorithms of finding out which of the $2^n$ initial configurations of width $n$ evolve to a specific configuration is also introduced. In Section 3, the development of microscopic computers and sub-microscopic computers for cellular automata is described. In Section 4, the plan of finding out which of the $2^n$ initial configurations of width $n$ evolve to a specific configuration in one-dimensional cellular automata is proposed. In Section 5, the time complexity and the space complexity of the proposed quantum algorithm are given. In Section 6, a brief conclusion is given.

2. ABSTRACTED PROBLEMS AND MOTIVATION

It is supposed that a system has $P = 2^n$ initial configurations of width $n$ that are labeled as $Q_0$, $Q_1$, $Q_2$, …, $Q_{P-1}$. These $2^n$ initial configurations of width $n$ are represented as $n$ bit strings. Let there be a unique initial configuration of width $n$, say $Q_k$ for $0 \leq k \leq 2^n - 1$, that satisfies the condition $F(Q_k) = 1$, whereas for all other initial configurations of width $n$, $Q_v$, for $0 \leq v \leq 2^n - 1$ and $v \neq k$, $F(Q_v) = 0$. This is to say that the unique initial configuration of width $n$, $Q_k$, with $F(Q_k) = 1$ evolves to a specific configuration, and for $0 \leq v \leq 2^n - 1$ and $v \neq k$, all other initial configurations of width $n$, $Q_v$, with $F(Q_v) = 0$ do not evolve to a specific configuration. The problem is to identify the unique initial configuration of width $n$, $Q_k$, which evolves to a specific configuration. Our motivation for writing the article is to ask how to design quantum circuits to identify $Q_k$ and how to complete amplitude amplification of the unique answer by means of Shor's quantum order-finding algorithm.

3. THE DEVELOPMENT OF MICROSCOPIC COMPUYERS AND SUB-MICROSCOPIC COMPUYERS FOR CELLULAR AUTOMATA

In [Neumann and Ulam 1966], cellular automata were originally introduced by von Neumann and Ulam as a system of biological self-reproduction. In [Wolfram 1982; Wolfram 1983; Wolfram 2002], a detailed analysis is given of elementary cellular automata (one-dimensional cellular automata) containing a sequence of sites with values 0 or 1 on a line, with each site evolving deterministically in discrete time steps in light of definite rules involving the values of its nearest neighbors. In [Gardner 1970], a two-state, two-dimensional cellular automaton (named Game of Life) was invented by Conway and was popularized by Gardner.

In [Nagel and Schreckenberg 1992], car traffic flow was originally conducted by Nagel and Schreckenberg, who developed a stochastic cellular automaton to simulate single-lane highway traffic. In [Benjamin et al.1996], for studying car traffic flow, another model with the addition of a 'slow-to-start' rule was developed by Benjamin, Johnson, and Hui. In [Rickert 1996 et al.], for highways have two lanes or more, Rickert et al. used cellular automata to check extra space in order to get the realistic behaviors of laminar to start-stop traffic flow. In [Wagner et al. 1997], Wagner et al. used cellular automata to design a two-lane simulation that accounts for a faster left lane which is to be used for passing. In [Esser and Schreckenberg 1997], based on cellular automata, Esser and Schreckenberg designed a complete simulation tool for urban traffic that accounts for realistic traffic light intersections, priority rules, parking capacities, and public transport circulation.

In [Cunha et al. 2005], Cunha et al. made reference to features of wireless sensor networks such as low computational power, low bandwidth capacity, limited energy supply, and high cost of real wireless sensor networks (hence the need for simulation), as reasons for why the applicability of cellular automata to simulate some aspects of sensor networks should be verified. In [Kwak et al. 2008], Kwak et al. designed a self organizing and energy efficient intrusion detection sensor system based on concepts from cellular automata. In [Frank and Romer 2004], based on an asynchronous cellular automaton, Frank and Romer proposed a robust and efficient generic role assignment scheme for wireless sensor networks. In [Subrata and Zomaya 2003], Subrata and Zomaya used cellular automata combined with a genetic algorithm to create an evolving parallel reporting cells planning algorithm for addressing the location tracking/management problem. In [Kirkpatrick and Scoy 2004], Kirkpatrick and Van Scoy investigated the use of cellular automata to model message broadcasting in highly mobile ad hoc networks.

In [Watrous 1995], Watrous proposed the first formal model of quantum cellular automata. In [Meyer 1996], Meyer presented another model of quantum cellular automata as a means of simulating quantum lattice gases. In [Tougaw and Lent 1994], Tougaw and Lent proposed a proposal for implementing *classical* cellular automata by systems designed with quantum dots has been proposed under the name "quantum cellular automata", as a replacement for classical computation using CMOS technology. In [Bhanja and Sarkar 2006], Bhanja and Sarkar introduced probabilistic modeling of quantum cellular automata circuits by means of using Bayesian networks. In [Srivastava and Bhanja 2007], Srivastava and Bhanja proposed probabilistic macro models for quantum circuits of cellular automata based on conditional probability characterization, defined over the output states given the input states. In [Henderson et al. 2004], Henderson et al. proposed to incorporate standard CMOS design process methodologies into logic design process of the quantum cellular automata. In [Graunke et al. 2005], Graunke et al. implemented a crossbar network using quantum-dot cellular automata.

## 4. PLAN OF SOLVING THE BACKTRACKING OF ONE-DIMENSIONAL CELLULAR AUTOMATA

In this section, we introduce the plan of finding out which of the $2^n$ initial configurations of width $n$ evolve to a specific configuration in one-dimensional cellular automata, and the quantum algorithm of solving the problem is also proposed.

### 4.1. INTRODUCTION OF ONE-DIMENSIONAL CELLULAR AUTOMATA

In [Wolfram 1983], cellular automata are mathematical idealizations of physical systems in which space and time are discrete, and physical quantities take on a finite set of discrete values. A cellular automaton contains a regular uniform lattice (or "array"), usually infinite in extent, with a discrete variable at each site ("cell") in [Wolfram 1983]. The state of a cellular automaton is completely specified by the values of the variables at each site in [Wolfram 1983]. A cellular automaton evolves in discrete time steps, with the value of the variable at one site being affected by the values of variables at sites in its "neighborhood" on the previous time step in [Wolfram 1983]. The neighborhood of a site is typically taken to be the site itself and all immediately adjacent sites. In [Wolfram 1983], the variables at each site are updated simultaneously ("synchronously"), based on the values of the variables in their neighborhood at the preceding time step, and in light of a definite set of "local rules." **Definition 4-1** cited from [Wolfram 1983] is used to explain one-dimensional cellular automata.

**Definition 4-1**: One-dimensional cellular automata contain a sequence of sites with values 0 or 1 on a line, with each site evolving deterministically in discrete time steps according to definite rules involving the values of its nearest neighbors.

Consider a one-dimensional cellular automaton in which its initial configuration of width eleven is 00000100000. This is to say that in the one-dimensional cellular automaton, the value of the sixth cell is 1, and the values of other cells are all 0. An important feature of cellular automata is that their behavior can readily be presented in a visual way in [Wolfram 2002]. It is assumed that if the value of one cell is 1, then the cell is colored as black. Also it is supposed that if the value of one cell is 0, then the cell is colored as white. Therefore, the initial configuration of the one-dimensional cellular automaton is shown in Figure 4-1 in [Wolfram 2002].

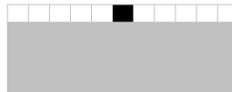

Figure 4-1: the initial configuration of the one-dimensional cellular automaton with a line of cells, 00000100000.

At every step there is then a definite rule that determines the status of a given cell from the status of that cell and its immediate left and right neighbors on the step before in [Wolfram 2002]. For the one-dimensional cellular automaton in Figure 4-1, a representation of the rule that is numbered as cellular automaton rule 254 in [Wolfram 2002] is shown in Figure 4-2. In Figure 4-2, the top row in each box gives one of the possible combinations of colors for a cell and its immediate neighbors. The bottom row then specifies what color the center cell should be on the next step in each of these cases. The evolving rule specifies that a cell should be black in all cases where it or either of its neighbors was black on the step before.

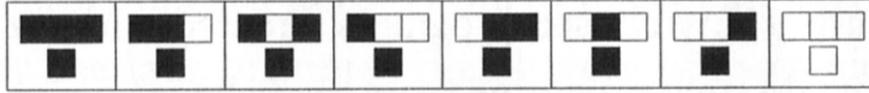

Figure 4-2: the evolving rule of the one-dimensional cellular automaton in Figure 4-1.

Based on the initial configuration in Figure 4-1 and the evolving rule in Figure 4-2, what the one-dimensional cellular automaton does over the course of five steps is shown in Figure 4-3 in [Wolfram 2002]. In Figure 4-3, a visual representation of the behavior is the evolving processing of the one-dimensional cellular automaton, and each row of cells corresponds to one evolving step. On the execution of the first evolving step in Figure 4-3, it is used to initialize the configuration of the one-dimensional cellular automaton. The cell in the center in the first row of cells is black and all other cells are white. Next, on the execution of the second evolving step, according to the seventh rule, the sixth rule and the fourth rule in Figure 4-2, the fifth cell, the sixth cell and the seventh cell in the second row of cells in Figure 4-3 are made black and other cells in the second row of cells in Figure 4-3 are made white. Next, on the execution of each successive evolving step, a particular cell is made black whenever it or either of its neighbors was black on the evolving step before. The evolving result of the one-dimensional cellular automaton leads to a simple expanding pattern uniformly filled with black in Figure 4-3.

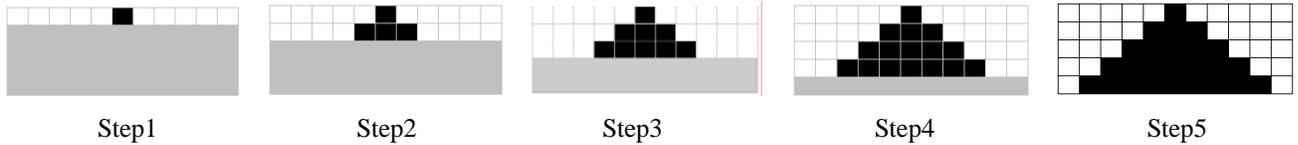

Step1  Step2  Step3  Step4  Step5

Figure 4-3: the evolving result of the one-dimensional cellular automaton in Figure 4-1.

### 4.2. INTRODUCTION OF THE BACKTRACKING OF ONE-DIMENSIONAL CELLULAR AUTOMATA AND A ONE-WAY FUNCTION

In Figure 4-3, the evolving result of the *fifth* Step in the one-dimensional cellular automaton is 01111111110. The backtracking of the one-dimensional cellular automaton in Figure 4-3 is to find out which initial configuration with eleven cells evolves to a specific configuration, 01111111110. It is very clear from Figure 4-3 that the answer (the initial configuration) is 00000100000. A function, $\Omega: \{c|0 \leq c \leq 2^n - 1\} \to \{d|0 \leq d \leq 2^m - 1\}$, is called a one-way function if figuring out $\Omega(c)$ for all $c$ can be completed fast but the opposite direction i.e. computing $c$ from any $d = \Omega(c)$ is demonstrated to be very hard in [Diffie and Hellman 1976; Imre and Balazs 2005]. The following lemma is used to show that a one-dimensional cellular automaton is a one-way function.

**Lemma 4-1:** A one-dimensional cellular automaton is a one-way function.

**Proof:**

**From Definition 4-1**, a one-dimensional cellular automaton consists of a sequence of sites with values 0 or 1 on a line, with each site evolving deterministically in discrete time steps according to definite rules involving the values of its nearest neighbors. It is assumed that the one-dimensional cellular automaton has $n$ cells on a line, where the value of each cell is 0 (white) or 1(black). The first initial configuration is a decimal value 0 with $n$ bits, the second initial configuration is a decimal value 1 with $n$ bits, and so on with that the last initial configuration is a decimal value $2^n - 1$ with $n$ bits. Also it is supposed that for each initial configuration evolving deterministically in discrete time steps according to definite rules involving the values of its nearest neighbors is a function, $\varepsilon: \{u|0 \leq u \leq 2^n - 1\} \to \{v|0 \leq v \leq 2^m - 1\}$. It is indicated from [Wolfram 2002] that computing $\varepsilon(u)$ for all $u$ can be completed efficiently (fast). This is to say that the last configuration of each initial configuration can be obtained fast. But the opposite direction i.e. figuring out $u$ from any $v = \varepsilon(u)$ is shown to be very hard in [Wolfram 2002]. This implies that it is very hard to find the corresponding initial configuration from a specific configuration. Therefore, it is inferred at once that a one-dimensional cellular automaton is a one-way function. ∎

### 4.3. ABSTRACTED DESCRIPTION FOR THE BACKTRACKING OF ONE-DIMENSIONAL CELLULAR

AUTOMATA AND A ONE-WAY FUNCTION IN A SYSTEM WITH $2^n$ STATES

From **Lemma 4-1**, a one-dimensional cellular automaton is a one-way function. Therefore, solving the backtracking of a one-dimensional cellular automaton is equivalent to solve the backtracking of a one-way function. It is assumed that for a one-dimensional cellular automaton with $n$ cells its domain, $\{u|0 \leq u \leq 2^n - 1\}$, is a set of all of the initial configurations. Also it is supposed that for the one-dimensional cellular automaton with $n$ cells its range, $\{v|0 \leq v \leq 2^m - 1\}$, is a set of all of the evolved configurations, and for each initial configuration evolving deterministically in discrete time steps according to definite rules involving the values of its nearest neighbors is a function, $\varepsilon\colon \{u|0 \leq u \leq 2^n - 1\} \rightarrow \{v|0 \leq v \leq 2^m - 1\}$. In a one-dimensional cellular automaton with $n$ cells, for a specific configuration $v = \varepsilon(u)$, the backtracking is to how to determine the corresponding unique initial configuration, $u$. This is equivalent to that in a one-way function, $\varepsilon\colon \{u|0 \leq u \leq 2^n - 1\} \rightarrow \{v|0 \leq v \leq 2^m - 1\}$, for given any $v = \varepsilon(u)$, the backtracking also is to how to figure out the corresponding unique solution, $u$.

It is assumed that a system has $Q = 2^n$ states which are labeled as $\theta_0, \theta_1, \theta_2, \ldots, \theta_{Q-1}$. It is also supposed that $\{u|0 \leq u \leq 2^n - 1\}$ can be encoded as those $2^n$ states of $n$ bit strings in the system. For example, state $\theta_0$ encodes a decimal value 0 with $n$ bits in the domain, state $\theta_1$ encodes a decimal value 1 with $n$ bits in the domain, and so on with that state $\theta_{Q-1}$ encodes a decimal value $(2^n - 1)$ with $n$ bits in the domain. It is supposed that one Boolean function $F(\theta_k)$ for $0 \leq k \leq 2^n - 1$ is applied to find its initial configuration, $u$, for a specific configuration $v = \varepsilon(u)$. If one is returned from $F(\theta_k)$, then state $\theta_k$ encodes a decimal value, $u$, that is the initial configuration of a specific configuration. This indicates that there is a unique state, say $\theta_k$ for $0 \leq k \leq 2^n - 1$, that satisfies the condition $F(\theta_k) = 1$, whereas for all other states $\theta_a$ for $0 \leq a \leq 2^n - 1$ and $a \neq k$, $F(\theta_a) = 0$ (it is supposed that for any state $\theta_b$ for $0 \leq b \leq 2^n - 1$, the condition $F(\theta_b)$ can be evaluated by means of polynomial quantum gates in [Grover 1994]). Thus, finding out which of the $2^n$ initial configurations of width $n$ evolve to a specific configuration is to label the state $\theta_k$.

4.4. INTRODUCTION OF ELEMENTARY NUMBER-THEORETIC NOTIONS

It is supposed that the set $Z = \{\ldots, -2, -1, 0, 1, 2, \ldots\}$ of integers and the set $\varpi = \{0, 1, 2, \ldots\}$ of natural numbers. The notation $d \mid a$ (read "$d$ divides $a$") means that $a = b \times d$ for some integer $b$. If $d \mid a$, then $a$ is a multiple of $d$, and $d$ is a divisor of $a$. For an integer $a > 1$, if its divisors only are, respectively, 1 and $a$, then it is a prime. For an integer $a > 1$, if it is not a prime, then it is a composite. The integer 1 is called as a unit, and it is neither prime nor composite. Similarly, the integer 0 and negative integers are neither prime nor composite. For any integer $a$ and any positive integer $N$, the value $a \bmod N$ is the remainder of the quotient $a/N$: $a \bmod N = a - N \times \lfloor a/N \rfloor$, where $\lfloor a/N \rfloor$ is to compute the biggest integer that is less than or equal to $a/N$.

In [Cormen et al. 2009], for any integer $a$ and any positive integer $N$, there exist unique integers $b$ and $c$ such that $0 \leq c < N$ and $a = b \times N + c$. The value $b = \lfloor a/N \rfloor$ is the quotient of the division. The value $c = a \bmod N$ is the remainder of the division. If any two integers, $a$ and $e$, have the same remainder when divided by $N$, then we write $a \equiv e \pmod{N}$ and say that $a$ is equivalent to $e$, modulo $N$. The equivalence class modulo $N$ containing an integer $e$ is: $[e]_N = \{e + q \times N : q \in Z\}$. Writing $a \in [e]_N$ is the same as writing $a \equiv e \pmod{N}$. If $d$ is a divisor of $a$ and $d$ is also a divisor of $e$, then $d$ is a common divisor of $a$ and $e$. The greatest common divisor of any two integers $a$ and $e$, not both zero, is the largest of the common divisor of $a$ and $e$; it is denoted as $\gcd(a, e)$.

4.5. INTRODUCTION OF A FINITE ABELIAN GROUP FOR THE BACKTRACKING OF ONE-DIMENSIONAL CELLULAR AUTOMATA IN A SYSTEM WITH $2^n$ PAIRS $(k, F(k), M(k))$

It is assumed that a system has $2^n$ pairs $(k, F(k), M(k))$ for $0 \leq k \leq 2^n - 1$ and $k$ encoding state $\theta_k$, where $F(k)$ is one Boolean function that is applied to find its initial configuration, $k$, for a specific configuration $v = \varepsilon(k)$ for $0 \leq k \leq 2^n - 1$, $F(k) \in \{0, 1\}$, and $G(k) = A^k \bmod N$ for that the greatest common divisor of $A$ and $N$ is equal to one. A more formal model for labeling the pair $(k, F(k) = 1, M(k))$ in the system, which we now give, is best described within the framework of group theory. Hence, D**efinition 4-1** through **Definition 4-3** cited from [Imre and Balazs 2005; Cormen et al. 2009] are employed to denote a group, an abelian group, and a finite group.

**Definition 4-1:** A group $(S, @)$ is a set $S$ together with a binary operation $@$ defined on $S$ for which the following properties hold:

1. **Closure:** For all $a, b \in S$, we have $a @ b \in S$.

2. **Identify:** There exists an element $e \in S$, called the *identity* of the group, such that $e @ a = a @ e = a$ for all $a \in S$.

3. **Associativity:** For all $a, b, c \in S$, we have $(a @ b) @ c = a @ (b @ c)$.

4. **Inverses:** For all $a \in S$, there exists a unique element $b \in S$, called the *inverse* of $a$, such that $a @ b = b @ a = e$.

**Definition 4-2:** A group $(S, @)$ is an abelian group if the group $(S, @)$ satisfies the commutative law $a @ b = b @ a$ for all $a, b \in S$.

**Definition 4-3:** A group $(S, @)$ is a finite group if the group $(S, @)$ satisfies $|S| < \infty$, where $|S|$ is the number of elements in $S$.

4.6. INTRODUCTION TO A RELATION OF DEGREE $a$ WITH $2^n$ ENTRIES FOR THE BACKTRACKING OF ONE-DIMENSIONAL CELLULAR AUTOMATA

The term *relation* is used here in its accepted mathematical sense. For any given sets $B_1, B_2, \ldots, B_a$, if $R$ is a set of $a$-tuples each of which has its first element from $B_1$, its second element from $B_2$, and so on with its last element from $B_a$, then it is a relation on these $a$ sets. This is to say that more concisely $R$ is a subset of the Cartesian product $B_1 \times B_2 \times \ldots \times B_a$. For $1 \leq j \leq a$, $S_j$ is referred to as the $j$th domain of $R$. In light of defined above, $R$ is said to have degree $a$. Relations of degree 1 are often called unary, degree 2 binary, degree 3 ternary, and degree $a$ $a$-ary. For the convenience of presentation and expository reasons, we shall frequently make use of an array representation of relations.

It is assumed that two positive integers $A$ and $N$ that are co-primes, i.e. $\gcd(A, N) = 1$, where $A < N$ and "$\gcd(A, N)$" is the greatest common divisor of $A$ and $N$. It is supposed that $\lceil \log_2(N \times N) \rceil$ is to figure out the smallest integer greater than or equal to $\log_2(N \times N)$. The *order* of $A$ in modulo $N$ is denoted as the least natural number $r$ of $n = \lceil \log_2(N \times N) \rceil$ bits such that $A^r \bmod N = 1$ and it is easy to see that $1 < r < N$ and $0 \leq r \leq 2^n - 1$. It is assumed that a system has $2^n$ initial configurations in which it includes the first function $F$: $\{k | 0 \leq k \leq 2^n - 1\} \to \{0, 1\}$, and the second function $M$: $\{k | 0 \leq k \leq 2^n - 1\} \to \{A^k \bmod N\}$. For the two functions $F$ and $M$, $\{k | 0 \leq k \leq 2^n - 1\}$ is their common domain. The range of $F$ is $\{0, 1\}$, and the range of $G$ is $\{A^k \bmod N / 0 \leq k \leq 2^n - 1\}$. In a one-dimensional cellular automaton with $n$ cells, its evolved function is $\varepsilon$: $\{k | 0 \leq k \leq 2^n - 1\} \to \{v | 0 \leq v \leq 2^m - 1\}$, where $\{k | 0 \leq k \leq 2^n - 1\}$ is a set of all of the initial configurations and $\{v | 0 \leq v \leq 2^m - 1\}$ is a set of all of the evolved configurations. If the first function $F$ finds the corresponding initial configuration $k$ for a specific configuration $v = \varepsilon(k)$, then $F(k) \in \{1\}$. Otherwise, $F(k) \in \{0\}$.

In Figure 4-4, a relation of degree 2, $R$, that is based on an array representation of relations is used to represent those results of implementing the two functions $F(k)$ and $M(k)$ for $0 \leq k \leq 2^n - 1$ in the system above. In Figure 4-4, the relation $R$ contains $2^n$ rows and each row consists of two columns. For the two functions $F$ and $M$, their common domain, $\{k | 0 \leq k \leq 2^n - 1\}$, is regarded as an index in the relation $R$ in Figure 4-4. This is to say that for a one-dimensional cellular automaton with $n$ cells, each initial configuration is regarded as an index in the relation $R$ in Figure 4-4. For a specific configuration $v = \varepsilon(k)$ to $0 \leq k \leq 2^n - 1$, if the first function $F(k)$ finds the corresponding initial configuration $k$, then $F(k) \in \{1\}$. Otherwise, $F(k) \in \{0\}$. Therefore, those values generated by the first functions $F(k)$ are regarded as the domain of the first column in the relation $R$ in Figure 4-4. Similarly, those values produced by the second functions $M(k)$ for $0 \leq k \leq 2^n - 1$ are regarded as the domain of the second column in the relation $R$ in Figure 4-4. In the first column of the first row, its value is $F(0)$, in the second column of the first row, its value is $M(0) = A^0 \bmod N$, and so on with that in the first column of the last row its value is $F(2^n)$ and in the second column of the last row its value is $M(2^n) = A^{2^n} \bmod N$.

| $F(k)$ | $M(k)$ |
|---|---|
| $F(0)$ | $M(0) = A^0 \bmod N$ |

| ⋮ | ⋮ |
| F(256) | $M(256) = A^{256} \bmod N$ |
| ⋮ | ⋮ |
| $F(2^n)$ | $M(2^n) = A^{2^n} \bmod N$ |

Figure 4-4: A relation of degree 2, R.

## 4.7. PROOF OF A FINITE ABELIAN GROUP FOR A RELATION OF DEGREE 2 WITH $2^n$ ENTRIES FOR THE BACKTRACKING OF ONE-DIMENSIONAL CELLULAR AUTOMATA

It is supposed that a relation of degree 2, R, in Figure 4-4 can be represented as a set $S = \{(k, F(k), M(k)) \mid 0 \leq k \leq 2^n - 1, F(k) \in \{0, 1\}, \text{ and } M(k) \in \{A^k \bmod N\}\}$. This is to say that finding the corresponding initial configuration $k$ for a specific configuration $v = \varepsilon(k)$ in a one-dimensional cellular automaton with $n$ cells is to find the only element $(k, F(k), M(k))$ in $S$ such that $F(k) \in \{1\}$. Since $r$ satisfies $A^r \bmod N = 1$, the set $S$ can be rewritten as $\{(z \times r + x, F(z \times r + x), A^{z \times r + x} \bmod N) \mid 0 \leq x \leq r-1, Z_x = \left\lfloor \frac{2^n - 1 - x}{r} \right\rfloor, \text{ and } 0 \leq z \leq Z_x\} = \{(z \times r + x, F(z \times r + x), A^x \bmod N) \mid 0 \leq x \leq r-1, Z_x = \left\lfloor \frac{2^n - 1 - x}{r} \right\rfloor, \text{ and } 0 \leq z \leq Z_x\}$. All of the elements in $S$ can be partitioned into $r$ equivalence classes in light of their remainders of modular exponentiation, $A^{z \times r + x} \bmod N = A^x \bmod N$. This indicates that each initial configuration and its corresponding evolved configuration also can be partitioned into $r$ equivalence classes according to the same condition. In $S$, the first equivalence class is $[0]_{A, N} = \{(z \times r + 0, F(z \times r + 0), A^0 \bmod N) \mid Z_0 = \left\lfloor \frac{2^n - 1 - 0}{r} \right\rfloor, \text{ and } 0 \leq z \leq Z_0\}$, the second equivalence class is $[1]_{A, N} = \{(z \times r + 1, F(z \times r + 1), A^1 \bmod N) \mid Z_1 = \left\lfloor \frac{2^n - 1 - 1}{r} \right\rfloor, \text{ and } 0 \leq z \leq Z_1\}$, and so on with that the last equivalence class is $[r-1]_{A, N} = \{(z \times r + r - 1, F(z \times r + r - 1), A^{r-1} \bmod N) \mid Z_{r-1} = \left\lfloor \frac{2^n - 1 - (r-1)}{r} \right\rfloor, \text{ and } 0 \leq z \leq Z_{r-1}\}$. A binary operation $+_{A, N}$ defined on a set $S$ is $[T]_{A, N} +_{A, N} [U]_{A, N} = [T + U]_{A, N}$, where $[T]_{A, N} \in S$, $[U]_{A, N} \in S$, and $0 \leq T$ and $U \leq r-1$. **Lemma 4-2** is employed to demonstrate that the system $(S, +_{A, N})$ is a finite abelian group.

**Lemma 4-2:** The system $(S, +_{A, N})$ is a finite abelian group.

**Proof:**

For all $[T]_{A, N}$ and $[U]_{A, N} \in S$, we have $[T]_{A, N} +_{A, N} [U]_{A, N} = [T + U]_{A, N}$. Since $0 \leq T$ and $U \leq r - 1$, we get $0 \leq T + U \leq 2 \times r - 2$. If $0 \leq T + U \leq r - 1$, then we obtain $[T + U]_{A, N} \in S$. If $r \leq T + U \leq 2 \times r - 2$, then we get $[T + U]_{A, N} = [T_1 + U_1 + r]_{A, N} = [T_1 + U_1]_{A, N} \in S$. Thus, we demonstrate that the system $(S, +_{A, N})$ is closed.

For all $[T]_{A,N}$, $[U]_{A,N}$ and $[V]_{A,N} \in S$, we have $([T]_{A,N} +_{A,N} [U]_{A,N}) +_{A,N} [V]_{A,N} = [T+U]_{A,N}$ $+_{A,N} [V]_{A,N} = [(T+U)+V]_{A,N} = [T+(U+V)]_{A,N} = [T]_{A,N} +_{A,N} [U+V]_{A,N} = [T]_{A,N} +_{A,N}$ $([U]_{A,N} +_{A,N} [U]_{A,N})$. Hence, we show associativity of $+_{A,N}$.

For all $[T]_{A,N}$ and $[U]_{A,N} \in S$, we have $[T]_{A,N} +_{A,N} [U]_{A,N} = [T+U]_{A,N} = [U+T]_{A,N} =$ $[U]_{A,N} +_{A,N} [T]_{A,N}$. Therefore, we demonstrate commutativity of $+_{A,N}$.

For all $[T]_{A,N}$ and $[0]_{A,N} \in S$, we have $[T]_{A,N} +_{A,N} [0]_{A,N} = [T+0]_{A,N} = [0+T]_{A,N} =$ $[0]_{A,N} +_{A,N} [T]_{A,N} = [T]_{A,N}$. Thus, we demonstrate that the identity element of $(S, +_{A,N})$ is $[0]_{A,N}$.

For each $[T]_{A,N} \in S$, we have $[T]_{A,N} +_{A,N} [r-T]_{A,N} = [T+(r-T)]_{A,N} = [(r-T)+T]_{A,N} =$ $[r-T]_{A,N} +_{A,N} [T]_{A,N} = [r]_{A,N} = [0]_{A,N}$. Hence, we show that the inverse of an element $[T]_{A,N}$ is $[r-T]_{A,N}$. From the proofs above, it is at once inferred that the system $(S, +_{A,N})$ is a finite abelian group. ∎

## 4.8. CONSTRUCTING $2^n$ INITIAL CONFIGURATIONS OF WIDTH $n$ FOR A RELATION OF DEGREE 2 WITH $2^n$ ENTRIES IN THE BACKTRACKING OF ONE-DIMENSIONAL CELLULAR AUTOMATA

In one-dimensional cellular automata, finding out which of the $2^n$ initial configurations of width $n$ evolves to a specific configuration can be regarded as constructing a finite Abelian group for a relation of degree 2, $R$, with $2^n$ entries in Figure 4-4. The quantum circuit of encoding $R$ in Figure 4-4 includes three main parts. The first part is for establishing an empty relation $R$ in Figure 4-4 with $2^n$ entries, the second part is for inserting those $2^n$ data generated by the first function $F(k)$ for $0 \leq k \leq 2^n - 1$ into the first column of each entry in the relation $R$, and the third part is for inserting those $2^n$ data produced by the second function $M(k)$ for $0 \leq k \leq 2^n - 1$ into the second column of each entry in the relation $R$.

It is supposed that a quantum register of $n$ bits, $(\otimes_{h=n}^{1} |k_h\rangle)$, is applied to initialize one-dimensional cellular automata that have $2^n$ initial configurations of width $n$ in which each initial configuration of width $n$ is one index of an empty relation $R$ in Figure 4-4. The initial states for $(\otimes_{h=n}^{1} |k_h\rangle)$ are set to $(\otimes_{h=n}^{1} |k_h^0\rangle)$. One-dimensional cellular automata that have $2^n$ initial configurations of width $n$ can be initialized to the distribution: $(\frac{1}{\sqrt{2^n}}, \frac{1}{\sqrt{2^n}}, \frac{1}{\sqrt{2^n}}, ..., \frac{1}{\sqrt{2^n}})$, i.e. there is the same amplitude to be in each of the $2^n$ initial configurations. This distribution can be obtained by means of $n$ Hadamard gates operating the initial quantum state vector, $(\otimes_{h=n}^{1} |k_h^0\rangle)$, for establishing an empty relation $R$ in Figure 4-4 with $2^n$ entries. Thus, the following new quantum state vector of encoding the empty relation $R$ in Figure 4-4 with $2^n$ entries is obtained

$$|\Omega_{4-1}\rangle = (H^{\otimes n})(\otimes_{h=n}^{1} |k_h^0\rangle) = \frac{1}{\sqrt{2^n}} (\otimes_{h=n}^{1}(|k_h^0\rangle + |k_h^1\rangle)) = \frac{1}{\sqrt{2^n}} (\sum_{k=0}^{2^n-1} |k\rangle). \tag{4-1}$$

In the new quantum state vector $|\Omega_{4-1}\rangle$ in (4-1), each initial configuration has the same amplitude, $\frac{1}{\sqrt{2^n}}$. The first initial configuration of width $n$ ($|0\rangle$) is used to encode the index of the first entry in $R$, the second initial configuration of width $n$ ($|1\rangle$) is applied to encode the index of the second entry in $R$, and so on with that the last initial configuration of width $n$ ($\left|2^n - 1\right\rangle$) is employed to encode the index of the last entry in $R$.

### 4.9. QUANTUM CIRCUITS OF EVOLVED RULES OF $2^n$ INITIAL CONFIGURATIONS WITH WIDTH $n$ FOR A RELATION OF DEGREE 2 WITH $2^n$ ENTRIES IN THE BACKTRACKING OF ONE-DIMENSIONAL CELLULAR AUTOMATA

Based on evolved rules, each initial configuration of width $n$ in one-dimensional cellular automata evolves to its final status. In fact, one of the $2^n$ initial configurations of width $n$ evolves to a specific configuration. It is assumed that evolved rules in one-dimensional cellular automata can be implemented by manes of one Boolean function $F(k)$ for $0 \leq k \leq 2^n - 1$. If $F(k)$ is equal to one, then state $k$ is the initial configuration of the specific configuration. Otherwise, state $k$ is not the initial configuration of the specific configuration. It is supposed that an auxiliary quantum register of $(A + 1)$ quantum bits, $(\otimes_{g=A}^{0} \left|\delta_g^{\,0}\right\rangle)$, is applied to record the evolved result to each initial configuration of width $n$. Also it is assumed that an auxiliary quantum register of one quantum bit, $(\left|0\right\rangle)$, is used to label the corresponding initial configuration of a specific configuration. Next, a unitary operator, $((U_F) \otimes (\otimes_{h=n}^{1} I_{2\times 2}) \otimes (I_{2\times 2}))$, is applied to implement the function of $F(k)$ for $0 \leq k \leq 2^n - 1$ and to operate the quantum state vector $(((\otimes_{g=A}^{0} \left|\delta_g^{\,0}\right\rangle) \otimes (\left|\Omega_{4-1}\right\rangle) \otimes (\left|0\right\rangle)))$ for inserting the evolved result of each initial configuration with width $n$ into the first column of each entry in a relation of degree, $R$, in Figure 4-4. Therefore, the following new quantum state vector is obtained

$$\left|\Omega_{4-2}\right\rangle = ((U_F) \otimes (\otimes_{h=n}^{1} I_{2\times 2}) \otimes (I_{2\times 2})) \,((\otimes_{g=A}^{0} \left|\delta_g^{\,0}\right\rangle) \otimes (\left|\Omega_{4-1}\right\rangle) \otimes (\left|0\right\rangle)) = (\frac{1}{\sqrt{2^n}})$$

$$(\sum_{k=0}^{2^n-1}(\otimes_{g=A}^{0}\left|\delta_g\right\rangle) \otimes (\left|k\right\rangle)) \otimes (\left|0\right\rangle). \tag{4-2}$$

Next, based on $\left|\delta_A\right\rangle$ that is a control bit and $\left|0\right\rangle$ that is a target bit, a unitary gate, $(((\otimes_{g=A}^{0} I_{2\times 2}) \otimes (\otimes_{h=n}^{1} I_{2\times 2}) \otimes (\left|\delta_A \oplus 0\right\rangle)))$, is used to operate the quantum state vector $(\left|\Omega_{4-2}\right\rangle)$ in (4-2). Thus, the following new quantum state vector is obtained

$$\left|\Omega_{4-3}\right\rangle = ((\otimes_{g=A}^{0} I_{2\times 2}) \otimes (\otimes_{h=n}^{1} I_{2\times 2}) \otimes (\left|\delta_A \oplus 0\right\rangle)) \,((\left|\Omega_{4-2}\right\rangle)) = \frac{1}{\sqrt{2^n}} (\sum_{k=0}^{2^n-1}(\otimes_{g=A}^{0}\left|\delta_g\right\rangle) \otimes (\left|k\right\rangle)$$

$$\otimes (\left|\delta_A \oplus 0\right\rangle)) = \frac{1}{\sqrt{2^n}} ((\sum_{w=0}^{k-1}(\otimes_{g=A}^{0}\left|\delta_g\right\rangle) \otimes (\left|w\right\rangle) \otimes (\left|F(w) = \delta_A = 0\right\rangle)) + ((\otimes_{g=A}^{0}\left|\delta_g\right\rangle) \otimes (\left|k\right\rangle)$$

$$\otimes (\left|F(k) = \delta_A = 1\right\rangle)) + (\sum_{w=k+1}^{2^n-1}(\otimes_{g=A}^{0}\left|\delta_g\right\rangle) \otimes (\left|w\right\rangle) \otimes (\left|F(w) = \delta_A = 0\right\rangle))) = \frac{1}{\sqrt{2^n}}$$

$$(\sum_{k=0}^{2^n-1}(\otimes_{g=A}^{0}\left|\delta_g\right\rangle) \otimes (\left|k\right\rangle) \otimes (\left|F(k)\right\rangle)). \tag{4-3}$$

In other words, with the operations above, we label the unique $(k, F(k) = 1)$ pair and all other $(w, F(w) = 0)$ pairs.

### 4.10. INTRODUCTION OF AMPLITUDE AMPLIFICATION FOR THE CORRESPONDING INITIAL CONFIGURATION OF A SPECIFIC CONFIGURATION AMONG $2^n$ INITIAL CONFIGURATIONS WITH WIDTH $n$ TO A RELATION OF DEGREE 2 WITH $2^n$ ENTRIES IN THE BACKTRACKING OF ONE-DIMENSIONAL CELLULAR AUTOMATA

Since quantum operations are naturally reversible and the auxiliary quantum bits can be restored to their initial states by means of reversing the operation in (4-2), the following quantum state vector is obtained

$$|\Omega_{4-4}\rangle = (\otimes_{g=A}^{0}|\delta_g\rangle) \otimes (\frac{1}{\sqrt{2^n}} (\sum_{k=0}^{2^n-1}(|k\rangle) \otimes (|F(k)\rangle))). \tag{4-4}$$

The new quantum state vector $|\Omega_{4-4}\rangle$ in (4-4) is actually made by the two subsystems. The first subsystem is

$$|\Omega_{4-5}\rangle = (\otimes_{g=A}^{0}|\delta_g\rangle). \tag{4-5}$$

The second subsystem is

$$|\Omega_{4-6}\rangle = \frac{1}{\sqrt{2^n}} (\sum_{k=0}^{2^n-1}(|k\rangle) \otimes (|F(k)\rangle)). \tag{4-6}$$

In the quantum state vector $|\Omega_{4-6}\rangle$ in (4-6), state $(|0\rangle \otimes |F(0)\rangle)$ encodes the value of the first column in the first row, state $(|1\rangle \otimes |F(1)\rangle)$ encodes the value of the first column in the second row, and so on with that state $(|2^n - 1\rangle \otimes |F(2^n - 1)\rangle)$ encodes the value of the first column in the last row. This is to say that the value of the first column in each row in $R$ in Figure 4-4 is inserted into the quantum state vector $|\Omega_{4-6}\rangle$ in (4-6). Although in the quantum state vector $|\Omega_{4-6}\rangle$ in (4-6) the unique answer with $(k, F(k) = 1)$ has been labeled, however the amplitude or probability for finding the unique answer with $(k, F(k) = 1)$ will decrease exponentially.

4.11. QUANTUM CIRCUITS OF CONSTRUCTING THE VALUE OF THE SECOND COLUMN IN A RELATION OF DEGREE 2 WITH $2^n$ ENTRIES IN A FINITE ABELIAN GROUP

For increasing exponentially the amplitude of the corresponding initial configuration of a specific configuration, it is supposed that $(|1\rangle \otimes (\otimes_{a=t-1}^{1}|0\rangle))$ is applied to store all the possible $A^k$ mod $N$ for $0 \leq k \leq 2^n - 1$, where $t = \lceil \log_2 N \rceil$, and it is also assumed that a unitary gate, $U_{SOFC}$, in Shor's quantum order-finding algorithm in [Shor 1994] is the corresponding quantum circuit of producing $A^k$ mod $N$ for $0 \leq k \leq 2^n - 1$. Therefore, the unitary operator, $U_{SOFC} = (\otimes_{h=n}^{1} I_{2\times 2}) \otimes (I_{2\times 2}) \otimes (|A^k \bmod N\rangle)$, is used to implement the function of $M(k)$ for $0 \leq k \leq 2^n - 1$ and to operate $((|\Omega_{4-6}\rangle) \otimes ((|1\rangle) \otimes (\otimes_{a=t-1}^{1}|0\rangle)))$ for inserting $2^n$ data into the second column of each entry in $R$ in Figure 4-4. So, the following quantum state vector is obtained

$$|\Omega_{4-7}\rangle = ((\otimes_{h=n}^{1} I_{2\times 2}) \otimes (I_{2\times 2}) \otimes (|A^k \bmod N\rangle)) ((|\Omega_{4-6}\rangle) \otimes ((|1\rangle) \otimes (\otimes_{a=t-1}^{1}|0\rangle))) = \frac{1}{\sqrt{2^n}}$$

$$(\sum_{k=0}^{2^n-1}|k\rangle \otimes |F(k)\rangle \otimes |A^k \bmod N\rangle) = \frac{1}{\sqrt{2^n}} (\sum_{k=0}^{r-1}(\sum_{z=0}^{Z_k}|z \times r + k\rangle |F(z \times r + k)\rangle) |A^k \bmod N\rangle), \tag{4-7}$$

where the least natural number $r$ of $n = \lceil \log_2(N \times N) \rceil$ bits satisfies $A^r \bmod N = 1$, $r$ is called the *order* of $A$ in modulo $N$, and $Z_k = \lfloor \frac{2^n - 1 - k}{r} \rfloor$. In the quantum state vector $|\Omega_{4-7}\rangle$ in (4-7), state $(|0\rangle |F(0)\rangle |A^0 \bmod N\rangle)$ encodes the value of each column in the first row, state $(|1\rangle |F(1)\rangle |A^1 \bmod N\rangle)$ encodes the value of each column in the second row, and so on with that state $(|2^n-1\rangle |F(2^n-1)\rangle |A^{2^n-1} \bmod N\rangle)$ encodes the value of each column in the last row. This indicates that the value of each column in each row in $R$ in Figure 4-4 is inserted into the quantum state vector $|\Omega_{4-7}\rangle$ in (4-7). Similarly, this also implies that based on the value of the second column all of the entries in $R$ in Figure 4-4 are classified as $r$ classes with that the $k$th class for $0 \leq k \leq r-1$ has the same value of the second column, $(|A^k \bmod N\rangle)$, and also has $Z_k = \left\lfloor \dfrac{2^n-1-k}{r} \right\rfloor$ entries.

## 4.12. QUANTUM CIRCUITS OF IMPLEMENTING AMPLITUDE AMPLIFICATION FOR THE CORRESPONDING INITIAL CONFIGURATION OF A SPECIFIC CONFIGURATION TO A RELATION OF DEGREE 2 WITH $2^n$ ENTRIES IN A FINITE ABELIAN GROUP

Because the amplitude in each entry in $R$ in Figure 4-4 is $\dfrac{1}{\sqrt{2^n}}$, after measuring one entry in $R$ in Figure 4-4, the success probability of seeing it may be exponentially small. Hence, for increasing the probability amplitude of each entry in $R$ in Figure 4-4, a unitary gate **IQFT** in Shor's quantum order-finding algorithm in [Shor 1994] that is the inverse of the quantum Fourier transform is applied to operate the first register in $(|\Omega_{4-7}\rangle)$, and the following quantum state vector is obtained

$$|\Omega_{4-8}\rangle = ((\mathbf{IQFT}) \otimes (|I_{2\times2}\rangle) \otimes ((|I_{2\times2}\rangle) \otimes (\otimes_{a=t-1}^{1} |I_{2\times2}\rangle))) (|\Omega_{4-7}\rangle) =$$

$$(\sum_{i=0}^{2^n-1}\sum_{k=0}^{r-1}(\underbrace{\sum_{z=0}^{Z_k} \frac{e^{-\sqrt{-1}\times\frac{2\pi}{2^n}\times i \times (z\times r + k)}}{2^n}}_{\varphi_{ik}}) |i\rangle |F(i)\rangle |A^k \bmod N\rangle). \qquad (4\text{-}8)$$

In the new quantum state vector $|\Omega_{4-8}\rangle$ in (4-8), for an index $|i\rangle$ its amplitude is

$$\sum_{k=0}^{r-1}(\underbrace{\sum_{z=0}^{Z_k} \frac{e^{-\sqrt{-1}\times\frac{2\pi}{2^n}\times i \times (z\times r + k)}}{2^n}}_{\varphi_{ik}}),$$ and for the value of the first column in the $i$th row, $|F(i)\rangle$, its amplitude is

$$\sum_{k=0}^{r-1}(\underbrace{\sum_{z=0}^{Z_k} \frac{e^{-\sqrt{-1}\times\frac{2\pi}{2^n}\times i \times (z\times r + k)}}{2^n}}_{\varphi_{ik}}).$$

## 4.13. A PROJECTION OF FINDING THE CORRESPONDING INITIAL CONFIGURATION OF A SPECIFIC CONFIGURATION FOR A RELATION OF DEGREE 2 WITH $2^n$ ENTRIES IN A FINITE ABELIAN GROUP

In the new quantum state vector $|\Omega_{4-8}\rangle$ in (4-8), it supports three different kinds of queries to obtain the required answer from a relation of degree 2, $R$, with $2^n$ entries in Figure 4-4. The first query is to find that the least natural number $r$ of $n = \lceil \log_2(N \times N) \rceil$ bits satisfies $A^r \bmod N = 1$. The second query is to obtain the value of the $i$th record in $R$ with $2^n$ entries in Figure 4-4. The third query is to find that the value of the first column in the $i$th row for $0 \leq i \leq 2^n - 1$ is equal to one ($F(i) = 1$). For the first query, if the measurement randomly selects an index (a computational basis vector) $|i\rangle$ in $R$ in Figure 4-4 so that $\frac{i}{2^n}$ is close to multiples of $\frac{1}{r}$ to complete continued fraction algorithms to extract $r$ from $i$, then the value of $r$ is obtained to be the order of $A$ in modulo $N$.

For the second query, if the measurement selects an index (a computational basis vector) $|i\rangle$ in $R$ in Figure 4-4, then the value of the $i$th record is obtained. For the third query, if a projection of $|F(i)\rangle$ onto $|1\rangle$ is made for the first column in $R$ in Figure 4-4, then it corresponds to the state collapse to a subspace relative to $|F(i)\rangle = |1\rangle$. If among $2^n$ indexes in $R$ in Figure 4-4 the only index satisfies the condition of the first function $F(i)$, then the only solution is obtained to be the index $i$ in $R$ in Figure 4-4. Otherwise, we obtain nothing in measurement. For the first query, the second query and the third query, if $\frac{i \times r}{2^n}$ is an integer, then the answer is obtained with a successful probability $\sum_{k=0}^{r-1}(\frac{Z_k+1}{2^n})^2$. Otherwise, the answer is obtained with a successful probability,

$$\sum_{k=0}^{r-1} \frac{1}{2^{2 \times n}} \times (\frac{\sin^2(\frac{\pi \times i \times r \times (Z_k+1)}{2^n})}{\sin^2(\frac{\pi \times i \times r}{2^n})}).$$

4.14. INTRODUCTION OF THE REASONS FOR APPLYING SHOR'S QUANTUM ORDER-FINDING ALGORITHM TO EXPONENTIALLY INCREASE THE AMPLITUDE OF THE CORRESPONDING INITIAL CONFIGURATION OF A SPECIFIC CONFIGURATION FOR A RELATION OF DEGREE 2 WITH $2^n$ ENTRIES IN A FINITE ABELIAN GROUP

In a one-dimensional cellular automaton with $n$ cells, its evolved function is $\varepsilon$: $\{k|0 \leq k \leq 2^n - 1\} \rightarrow \{v|0 \leq v \leq 2^m - 1\}$, where $\{k|0 \leq k \leq 2^n - 1\}$ is a set of all of the initial configurations and $\{v|0 \leq v \leq 2^m - 1\}$ is a set of all of the evolved configurations. Calculating the corresponding initial configuration $k$ from a specific configuration $v = \varepsilon(k)$ is a difficulty task on a digital computer. Similarly, it is also a difficulty task on a digital computer to determine that the *order* of $A$ in modulo $N$ is defined as the least natural number $r$ of $n = \lceil \log_2(N \times N) \rceil$ bits such that $A^r$ mod $N = 1$ and it is easy to see that $1 < r < N$ and $0 \leq r \leq 2^n - 1$. Because their domain is the same, solving the two problems can be reduced to construct a relation of degree 2, $R$, in Figure 4-4 and to complete the required queries on the relation, $R$. From **Lemma 4-2**, a relation of degree 2, $R$, in Figure 4-4 is proven to be a finite abelian group. Hence, the inverse of the quantum Fourier transforms is used amplify the amplitude of each element in a finite abelian group. This implies that the amplitude of each entry in a relation of degree 2, $R$, in Figure 4-4 is also amplified. Quantum circuits of inserting $2^n$ data into the second column of each entry and quantum circuits of amplifying the amplitude of each entry are actually implemented by Shor's quantum order-finding algorithm, so it is very clear that Shor's quantum order-finding algorithm can be used to amplify the amplitude of the corresponding initial configuration $k$ from a specific configuration $v = \varepsilon(k)$ in a one-dimensional cellular automaton with $n$ cells.

5. COMPLEXITY ASSESSMENT

The following lemmas are used to show the time and space complexity of the proposed quantum algorithm for solving the backtracking of one-dimensional cellular automata in a relation of degree 2 with $2^n$ entries in a finite Abelian group.

**Lemma 5-1:** The time complexity of the proposed quantum algorithm for solving the backtracking of one-dimensional cellular automata with $n$ cells in a relation of degree 2 with $2^n$ entries in a finite Abelian group is polynomial quantum gates, and the successful probability of finding the corresponding initial configuration for a specific configuration is the same as that of Shor's order-finding algorithm.

**Proof:**

It is shown from (4-1) through (4-6) that the number of unitary operators for labeling the corresponding initial configuration of a specific configuration in one-dimensional cellular automata in a relation of degree 2 with $2^n$ entries in a finite Abelian group is polynomial quantum gates. It is also very clear from [Shor 1994] that the number of unitary operators for implementing Shor's order-finding algorithm from (4-7) through (4-8) is also polynomial quantum gates. In light of the statements above, thus, it is at once inferred that the time complexity of the proposed quantum algorithm for solving the backtracking of one-dimensional cellular automata with $n$ cells in a relation of degree 2 with $2^n$ entries in a finite Abelian group is polynomial quantum gates, and the successful probability of finding the corresponding initial configuration for a specific configuration is the same as that of Shor's order-finding algorithm. ∎

**Lemma 5-2:** The space complexity of the proposed quantum algorithm for solving the backtracking of one-dimensional cellular automata with $n$ cells in a relation of degree 2 with $2^n$ entries in a finite Abelian group is $O(n + A + t + 2)$ quantum bits.

**Proof:**

From (4-1) through (4-8), a quantum register of $n$ quantum bits, $(\otimes_{h=n}^{1} |k_h^0\rangle)$, is used to initialize $2^n$ initial configurations of width $n$ in one-dimensional cellular automata, an auxiliary quantum register of $(A + 1)$ quantum bits, $(\otimes_{g=A}^{0} |\delta_g^0\rangle)$, is employed to store the evolved result of each initial configuration, an auxiliary quantum register of one quantum bit, $(|0\rangle)$, is used to label the answer, and an auxiliary quantum register of $t$ quantum bits, $((|1\rangle) \otimes (\otimes_{a=t-1}^{1} |0\rangle))$ is applied to exponentially increase the amplitude or probability of finding the answer. Therefore, it is at once derived that the space complexity of the proposed quantum algorithm for solving the backtracking of one-dimensional cellular automata is $O(n + A + t + 2)$ quantum bits in a relation of degree 2 with $2^n$ entries in a finite Abelian group. ∎

## 6. CONCLUSIONS

In [Imre and Balazs 2007], quantum counting algorithm for a classical engineering problem with $2^n$ possible solutions can be used to efficiently compute the number of real solutions, $\alpha$. If for the classical engineering problem $(\frac{\alpha}{2^n})$ is equal to $(\frac{1}{4})$, then Grover's algorithm in [Grover 1996] can take advantage of quantum mechanics to complete nanoscale computing with that the successful probability of measuring the answer is one and can obtain an exponential improvement compared with classical brute force search. Otherwise, Grover's algorithm only can obtain a quadratic improvement compared with classical brute force search, and the successful probability of measuring the answer is at least $\frac{1}{2}$. In [Shor 1994], Shor's order-finding algorithm for dealing with the problem of factoring integers can obtain an exponential improvement compared with classical methods. In [Boneh and Lipton 1995], for dealing with the discrete log problem over any group including Galois fields and elliptic curves, their proposed method can also obtain an exponential improvement compared with classical methods.

In [Wolfram 2002], the backtracking of one-dimensional cellular automata is to find out which of the $2^n$ possible initial configurations of width $n$ evolves to a specific configuration. From **Lemma 5-1**, the time complexity of the proposed quantum algorithm for solving the problem is polynomial quantum gates, the successful probability of finding the corresponding initial configuration is the same as that of Shor's order-finding algorithm, and from **Lemma 5-2**, the space complexity of the proposed quantum algorithm is $O(n + A + t + 2)$ quantum bits. This implies that for dealing with the backtracking of one-dimensional cellular automata, the proposed quantum algorithm gives an exponential improvement compared with classical methods.

# REFERENCE

Adleman L. 1994. Molecular computation of solutions to combinatorial problems. *Science*, 266: 1021-1024, November 11.

Benjamin S. C., Johnson N. F., and Hui P. M. 1996. Cellular automata models of traffic flow along a highway containing a junction. *Journal of Physics A: Mathematical and General*, 29(12): pp. 3119-3127.

Bhanja S. and Sarkar S. 2006. Probabilistic modeling of QCA circuits using bayesian networks. *IEEE Transactions on Nanotechnology*, Volune 5(6), pp. 657-670.

Boneh D. and Lipton R. J. 1995. Quantum cryptoanalysis of hidden linear functions. In Don Coppersmith, editor, *CRYPTO 95*, Lecture Notes in Computer Science, pp. 424-437.

Cormen T. H., Leiserson C. E., Rivest R. L., and Stein C. 2009. *Introduction to Algorithms*. The MIT Press, the third edition.

Cunha R. O., Silva A. P., Loreiro A. A. F., and Ruiz L. B. 2005. Simulating large wireless sensor networks using cellular automata. *Simulation Symposium*, 2005, *Proceedings*, 38th Annual, pp. 323-330, 4-6 April.

Diffie W. and Hellman M. 1976. New directions in cryptography. *IEEE Transaction on Information Theory*, Volume IT-22, pp. 644-654.

Esser J. and Schreckenberg M. 1997. Microscopic simulation of urban traffic based on cellular automata. *International Journal of Modern Physics C*, 8(5): pp. 1025-1036.

Frank C. and Romer K. 2004. Algorithms for generic role assignment in wireless sensor networks. *In 11th ACM SIGOPS European Workshop*, pp. 7-12, September.

Gardner M. 1970. Mathematical games: Conway's game of life. Scientific American, 223, pp. 120-123, October.

Graunke C., Wheeler D., Tougaw D. and Will J. D. 2005. Implementation of a crossbar network using quantum-dot cellular automata. *IEEE Transactions on Nanotechnology*, Volume 4, No. 4, pp. 1 – 6, July.

Grover L. K. 1996. A fast quantum mechanical algorithm for database search. *Proceedings of the twenty-eighth annual ACM symposium on Theory of computing*, pp. 212-219, May.

Henderson S., Johnson E., Janulis J. and Tougaw D. 2004. Incorporating standard CMOS design process methodologies into the QCA logic design process. *IEEE Transaction on Nanotechnology*, Volume 3, No. 1, pp. 2 – 9, March.

Imre S. and Balazs F. 2007. *Quantum Computation and Communications*: *An Engineering Approach*. John Wiley & Sons, Ltd, ISBN 0-470-86902-X.

Kirkpatrick M. and Scoy F. L. V. 2004. Using cellular automata to determine bounds for measuring the efficiency of broadcast algorithms in highly mobile ad hoc networks. *Lecture Notes in Computer Science*, Volume 3305, pp. 316-324.

Kwak K. J., Baryshnikov Y. M., and Coffman E. G. 2008. Cyclic cellular automata: a tool for self-organizing sleep scheduling in sensor networks. *International Conference on Information Processing in Sensor Networks*, pp. 535 – 536, 22-24 April.

Meyer D. 1996. From quantum cellular automata to quantum lattice gases. *Journal of Statistical Physics* 85, pp. 551–574.

Nagel K. and Schreckenberg M. 1992. A cellular automaton model for freeway traffic. *Journal de Physique I*, 2(12): pp. 2221-2229.

Rickert M., Nagel K., Schreckenberg M., and Latour A. 1996. Two lane traffic simulations using cellular automata. *Physica A*: *Statistical and Theoretical Physics*, 31(4): pp. 534-550.

Shor P. W. 1994. Algorithm for quantum computation: discrete logarithm and factoring algorithm. Proceedings of the 35th Annual IEEE Symposium on Foundation of Computer Science, pp.124-134, 1994.

Srivastava S. and Bhanja S 2007. Hierarchical probabilistic macro modeling for QCA circuits. *IEEE Transactions on Computers*,Volume 56(2), pp. 174-190.

Subrata R. and Zomaya A. Y. 2003. Evolving cellular automata for location management in mobile computing networks. I*EEE Transaction Parallel Distribution System*,14(1): pp. 13-26.

Tougaw P. and Lent C. 1994. Logical devices implemented using quantum cellular automata. *Journal of Applied Physics*, *Volume* 75, pp. 1818–1825.

Von Neumann J. 1966. The general and logical theory of automata. In J. von Neumann. "Collected works" (editor A. H. Taub), Volume 5, 288; von Neumann J., Theory of self-reproducing automata, (editor A.W. Burks), University of Illinois Press (1966); editor A.W. Burks, "Essays on cellular automata", University of Illinois Press (1970).

Wagner P., Nagel K., and Wolf D. E. 1997. Realistic multi-lane traffic rules for cellular automata. *Physica A: Statistical and Theoretical Physics*, 234(3-4): pp. 687-698.

Watrous J. 1995. On one-dimensional quantum cellular automata. *In The 36th Annual Symposium on Foundations of Computer Science*, pp. 528–537.

Wolfram S. 1982. Cellular automata as simple self-organizing systems. Caltech Preprint CALT-68-938.

Wolfram S. 1983. Statistical mechanics of cellular automata. Rev. Mod. Phys. 55, pp. 601–644.

Wolfram S. 2002. *A New Kind of Science*. Champaign, IL: Wolfram Media, Inc., **ISBN** 1-57955-008-8.